\def\kbf{{\bf k}}
\def\rbf{{\bf r}}
\def\ksum{\sum_{\kbf}}
\def\vec#1{\vert #1 \rangle}
\def\rsr{\langle r \rangle}

\documentstyle[aps,prl,epsf]{revtex}
\bibstyle{unsrt}
\title {OPERATOR METHOD IN SOLVING 
       NON-LINEAR EQUATIONS OF 
       THE HARTREE-FOCK TYPE}
\author{Le Anh Thu \thanks{Present address: 
               Max-Planck Institute for Physics of Complex Systems,
                N\"othnitzer Str. 38,
		D-01187 Dresden,
		Germany. E-mail: lathu@idefix.mpipks-dresden.mpg.de}$^1$ and		
	 L I Komarov$^2$}
\address{${}^1$ Institute of Physics, P.O. Box 429 Bo Ho, Hanoi 10000, Vietnam}
\address{${}^2$ Department of Physics, Belarussian State University, 
               4 Fr. Skariny av., Minsk 220050, Republic of Belarus}       
\date{\today}       
    
\begin{document}
\maketitle
\begin{abstract}
The operator method is used to construct the solutions of the problem of the 
polaron in the strong coupling limit and of the helium atom on the basis of the
Hartree-Fock equation. $E_0=-0.1085128052\alpha^2$ is obtained for the
polaron ground-state energy. Energies for $2s$- and $3s$-states are also calculated.
The other excited states are briefly discussed.
\end{abstract}

\section{Introduction}

The operator method (OM) of the approximate solution of the Schr\"odinger
equation was suggested in the works of Feranchuk and Komarov
[1-3]. In agreement with the results obtained in these and subsequent
works (see, for example, refs. [4-12]), the solution in the zeroth 
approximation of the OM gives quite a
simple and universal algorithm for obtaining an approximation, which is
uniformly applicable in a wide range of variation of parameters of the
Hamiltonian. Another advantage of the OM is connected with the possibility of
regularly calculating the correction to the zeroth approximation. Here, in
any order the needed calculations are reduced to a simple algebraic procedure
of expressing the product of excitation creation and annihilation operators
in the normal form, and this essentially simplified the use of computers.

As shown by the results of many works the OM is very effective for solving
various physical problems. However, the question of the application of this
method in solving non-linear equations, which are often met in atomic and solid
state physics, remains unclear at present.

In this work we investigate in two examples the possibility of using the OM 
for solving non-linear equations such as that of Hartree-Fock:

(i) in an example of the problem of a polaron in the regime of strong
electron-phonon coupling (section II);

(ii) in an example of the problem of a helium atom (section III).

It should be noted that such calculations have been carried out on the basis
of various methods. But here, we want to obtain the exact solution by a 
numerical method. Comparison of approximate solutions with the exact 
numerical solution enables us to estimate not only the accuracy of the 
approximate solutions, but also the usefulness of their further study.

\section{The operator method in the polaron problem}

The problem of a polaron in the strong coupling regime has become of interest since
Emin \cite{emin} proposed the mechanism of bipolaron Bose-Einstein
condensation to explain high-{$T_c$} superconductivity, and on the other hand,
the bipolaron is known to exist in this strong coupling regime \cite {smond2}.
The Hamiltonian of a non-relativistic particle (electron) interacting with the
quantized scalar field of lattice oscillations in an ionic crystal 
has the form
(see, for example, refs. [6,7,15,16] and reviews [17,18]): 
  $$\hat H=-{\triangle\over 2m}+\ksum \omega_{\kbf}c^+_{\kbf}c_{\kbf}
    +{g\over\sqrt{\Omega}}\ksum\biggl( A_{\kbf}e^{i\kbf\rbf}c_{\kbf}
    +A^*_{\kbf}e^{-i\kbf\rbf}c^+_{\kbf}\biggr). \eqno(1)$$
\noindent
Here, $c^+_{\kbf}$ and $c_{\kbf}$ are respectively the operators of creation
and annihilation of phonons with frequency $\omega_{\kbf}$ and momentum
$\kbf$, $A_k$ are the Fourier components of the source density and $\Omega$ 
is the volume of the system. In the theory of the so-called optical polaron,
the frequency of phonons is assumed to be independent of their wavevector.
Usually, it is assumed that 
  $$\omega_{\kbf}=\omega_0, \qquad 
    gA_{\kbf}=-{i\omega_0\over k}{(4\pi\alpha)^{1/2}
    \over (2m\omega_0)^{1/4}},$$
where $\alpha$ is a dimensionless electron-phonon coupling constant.

For the application of the OM it is conducive to cross to the four-dimensional
oscillator representation using the transformation \cite {k}
  $$\cases{ x_{\lambda} = \xi^*_s (\sigma_{\lambda})_{st} \xi_t \cr
               \chi     = \arg (\xi_1),\cr} $$	    
where $\sigma_{\lambda}\;(\lambda=1,2,3)$ are Pauli matrixes, $\xi_s (s=1,2)$ 
are regarded as spinor components. The physical meaning of ``extra''
variable $\chi$ has been treated in many works (see, for example, refs. [20,21]
and references therein).

Let us define operators
   $$  a_s(\omega)=\sqrt{\frac{\omega}{2}}\left(\xi_s+\frac{1}{\omega}
       \frac {\partial}{\partial \xi^*_s}\right), \qquad  
       b_s(\omega)=\sqrt{\frac{\omega}{2}} \left(\xi^*_s+\frac{1} {\omega}\frac
       {\partial}{\partial \xi_s}\right),$$ 
    $$  a^+_s(\omega)=\sqrt{\frac{\omega}{2}}\left(\xi^*_s-\frac{1}{\omega}
        \frac {\partial}{\partial \xi_s}\right), \qquad
       b^+_s(\omega)=\sqrt{\frac{\omega}{2}} \left(\xi_s - \frac{1} 
       {\omega}\frac {\partial}{\partial\xi^*_s}\right). \eqno(2)$$
Here, the positive parameter $\omega$ will be defined later.
Operators (2) satisfy the commutation relations
  $$ \lbrack a_s(\omega),a^+_t(\omega) \rbrack = \delta_{st}, \qquad
   \lbrack b_s(\omega),b^+_t(\omega) \rbrack = \delta_{st} $$
(we have written only non-zero commutators). The possibility of using the
algebraic method is conditioned by the fact that, all ``physical''
operators can be expressed through the following 15 operators
  $$ M=a_sb_s, \qquad M^{+}=a^+_sb^+_s , \qquad N=a^+_sa_s+b^+_sb_s ,\eqno(3)$$ 
  $$ m_{\lambda}=(\sigma_{\lambda})_{st}a_tb_s,  \qquad
   m^+_{\lambda}=(\sigma_{\lambda})_{st}a^+_sb^+_t, \qquad
  n^a_{\lambda}=(\sigma_{\lambda})_{st}a^+_sa_t, \qquad
  n^b_{\lambda}=(\sigma_{\lambda})_{st}b^+_tb_s,$$
which form a closed algebra $so(4,2)$ (see ref. \cite {k}). 
Here and henceforth,
we omit for brevity the parameter $\omega$ in expressions of the operators.
For further use we rewrite some operators into this representation as
follows 
  $$x_{\lambda}={1\over 2\omega}(m_{\lambda}^++m_{\lambda}+n^a_{\lambda}
                +n^b_{\lambda}), \qquad
  r=\xi_s^*\xi_s={1\over 2\omega}(M^++M+N+2),$$
  $$ r\triangle=-{\partial^2 \over \partial\xi_s^*\partial\xi_s}=
    {\omega \over 2}(2+N-M-M^+),$$
where the operators $N, M, M^+$ form the subalgebra $so(2,1)$ and satisfy
the following relations
  $$[M,M^+]=2+N, \qquad [M,2+N]=2M, \qquad [2+N,M^+]=2M^+. \eqno(4)$$
In the four-dimensional oscillator representation ($\xi$-space) the equation for
eigenvalue problem $\hat H \vec{\Psi} = E \vec{\Psi}$ has the form \cite{k}
  $$\hat L\vec{\Psi}\equiv r(\hat H-E)\vec{\Psi}=0, \eqno(5) $$
which, in terms of the operators (3), rather simplifies the algebraic
calculation of the matrix elements. It is easy to see that this representation is especially
useful for the bound polaron problem and leads to the equation without singular Coulomb term.
Further, we shall limit ourselves to case of the strong coupling regime, i.e.
when $\alpha\gg 1$. It is a well known fact that, in this limit, one can neglect
the quantum fluctuations of the phonon field (see \cite {13} and a recent discussion
in \cite {lieb}). 
After introducing the canonical transformations
  $$ c_{\kbf}=b_{\kbf}+u_{\kbf}, \qquad
     c_{\kbf}^+=b_{\kbf}^++u^*_{\kbf} \eqno(6) $$
and neglecting the quantum components, we then have
  $$\hat L=-{1\over 2} r\triangle+2^{3/4}\biggl({\pi\alpha\over\Omega}\biggr)
    ^{1/2}\ksum {1\over k}\biggl( re^{i\kbf\rbf}u_{\kbf}
    +re^{-i\kbf\rbf}u^*_{\kbf}\biggr)
    +r\biggl(\ksum u^*_{\kbf}u_{\kbf}-E\biggr) \eqno(7)$$
(here, all measurements are in the system of units whereby 
$m=\hbar=\omega_0=1$). The classical components of the field can be defined
from the condition
  $${\partial E\over\partial u_{\kbf}}=
    {\partial E\over\partial u^*_{\kbf}}=0. \eqno(8)$$
From equations (5), (7) and (8) we obtain
  $$u_{\kbf}=-2^{3/4}\biggl({\pi\alpha\over\Omega}\biggr)^{1\over 2}
    {1\over\rsr k}\langle re^{-i\kbf\rbf}\rangle. \eqno(9)$$
Here $\langle...\rangle$ represents the average 
$\langle\Psi|...|\Psi\rangle$, where $|\Psi\rangle$ is the polaron
eigenfunction. Substituting (9) into (7) and than integrating over
$\kbf$ we have
  $$\hat L=-{1\over 2}r\triangle -2^{1/2}\alpha\;{1\over\rsr}\int d^4\eta 
     \;{rr'|\Psi(\eta)|^2\over |\rbf -\rbf'|}
 +r\biggl({\alpha\over 2^{1/2}\rsr^2}\int d^4\eta\int d^4\zeta\;
   {r'r''\over |\rbf'-\rbf''|}\;|\Psi(\eta)|^2|\Psi(\zeta)|^2-E\biggr), 
   \eqno(10)$$
where
  $$x'_{\lambda}=\eta^*_s(\sigma_{\lambda})_{st}\eta_t, \qquad
    x''_{\lambda}=\zeta^*_s(\sigma_{\lambda})_{st}\zeta_t.$$
The emergence of the terms $\rsr$ and $\rsr^2$ is conditioned
by the changes in the condition of normalization of wavefunctions in the 
$\xi$-space. Further, we shall consider only $s$-states
of the polaron. The
general case will be treated at the end of this paper. Given the fact that
the eigenfunctions of $s$-states spherically symmetric, i.e. they depend only
on $r$, we can write $\Psi(\xi)\equiv\Psi(r)$. 
Averaging (10) over the angles and using the formula 
  $${1\over 2}\;\int^1_{-1}dx{1\over\sqrt{r^2+r'^2-2xrr'}}={1\over rr'}
    [r'\theta(r-r')+r\theta(r'-r)],$$
where the Heaviside function $\theta(x)$ vanishes for $x<0$, is $1/2$ for $x=0$ and unity
for $x>0$, we have 
  $$\Biggl(-{1\over 2}r\triangle -\sqrt 2\;\alpha\;{1\over\rsr}\int d^4\eta\;
    |\Psi(\eta)|^2[r'\theta(r-r')+r\theta(r'-r)]+ $$
  $$ +{\alpha\over \sqrt 2\rsr^2}r\int d^4\eta\int d^4\zeta\;
    |\Psi(\eta)|^2|\Psi(\zeta)|^2[r'\theta(r''-r')+r''\theta(r'-r'')]-rE
    \Biggr)\vec {\Psi(\xi)}=0.
    \eqno(11)$$
It is clear that equation (11) is an integro-differential equation of the
Hartree-Fock type.
This equation can be simplified using new units of energy and length,
such units are equal to the old ones multiplied respectively by $\alpha^2$
and $1/\alpha$. In this units the parameter $\alpha$ disappears in the last
equation. Therefore, we can further put $\alpha=1$.

Let assume that the polaron eigenfunction has the form
  $$\vec{\Psi}=\sum_n C_n\vec{n} \eqno(12) $$
where
  $$\vec{n}={1\over\sqrt{n!(n+1)!}}\Bigl(M^+\Bigr)^n\vec{0}.$$
It is clear that we need to calculate the matrix elements of the operators
constituting $\hat L$, such as
  $$\langle m_1,m_2|[r_2\theta(r_1-r_2)+r_1\theta(r_2-r_1)]
    \vec{n_1,n_2}\equiv W_{m_1 m_2;n_1 n_2}. \eqno(13)$$
To this effect we use the integral representation 
  $$r_2\theta(r_1-r_2)+r_1\theta(r_2-r_1)=\left ({1\over 2\pi}\right)^2
    \int\limits_{-\infty}^{+\infty} dq_1\int\limits_{-\infty}^{+\infty} dq_2
   \; e^{iq_1r_1+iq_2r_2}\varphi(q_1,q_2), \eqno(14)$$
where
  $$\varphi(q_1,q_2)=\int\limits_0^{\infty}dr_1\int\limits_0^{\infty}dr_2
    [r_2\theta(r_1-r_2)+r_1\theta(r_2-r_1)]e^{-iq_1r_1-iq_2r_2}.$$
The operator
  $$\exp(iqr)=\exp\left({iq\over 2\omega}(2+N+M+M^+)\right)$$
can be expressed in the normal form as follows (see, for example, \cite{k}, 
\cite {wilcox})
  $$\exp(iqr)={1\over (1-\mu)^2}\exp\biggl({1 \over 1-\mu}M^+
    \biggr)\exp\biggl(-N\ln(1-\mu)\biggr)
 \exp\biggl({1\over 1-\mu}M\biggr),$$
where, $\mu={iq\over 2\omega}$.
As a result, the calculation of matrix elements of the operator $\exp(iqr)$ 
using the algebra (4)
does not pose much difficulties. Here, we give only the result obtained
  $$\langle m|\exp(iqr)\vec{n}=\sqrt{n+1\over m+1}\sum^n_{s=0}C^{s}_n
    C^{s+1}_n{(\mu)^{m+n-2s}\over(1-\mu)^{m+n+2}}, \eqno(15)$$
where
  $$C^p_q={q!\over p!(q-p)!}.$$
Subtituting (15) into (13) and (14),
after integration over the variables $q_1,q_2,r_1,r_2$ we finally have the
formula 
  $$W_{m_1 m_2;n_1 n_2}={1\over 16\omega}(-1)^{m_1+m_2+n_1+n_2}
    \sqrt{(n_1+1)(n_2+1)\over (m_1+1)(m_2+1)}
    \sum^{n_1}_{s_1=0}C^{s_1}_{n_1}C^{s_1+1}_{m_1+1}$$
  $$\times\sum^{m_1+n_1-2s_1}_{t_1=0}(-1)^{t_1}C^{t_1}_{m_1+n_1-2s_1}
    \sum^{n_2}_{s_2=0}C^{s_2}_{n_2}C^{s_2+1}_{m_2+1}
    \sum^{m_2+n_2-2s_2}_{t_2=0}(-1)^{t_2}C^{t_2}_{m_2+n_2-2s_2}$$
  $$\times(2s_1+t_1+2s_2+t_2+3)(2s_1+t_1+2s_2+t_2+4)C^{2s_1+t_1+1}
    _{2s_1+t_1+2s_2+t_2+2} $$
  $$\times\left\{ \sum^{2s_1+t_1+1}_{u=0}{(-1)^uC^u_{2s_1+t_1+1}\over
     (2s_2+t_2+3+u)2^{2s_2+t_2+u}}+
    \sum^{2s_2+t_2+1}_{u=0}{(-1)^uC^u_{2s_2+t_2+1}\over
     (2s_1+t_1+3+u)2^{2s_1+t_1+u}}\right\}.\eqno(16)$$
Other matrix elements are calculated in the same way.

Let us now consider the zeroth approximation of the OM. The free parameter     
$\omega$ is chosen in such a way that the condition
  $${\partial E^{(0)}\over\partial\omega }= 0$$
is satisfied \cite{3}. Whence
  $$\omega_n^{(0)}={W_{nnnn}\over 2\sqrt{2}(n+1)^2},
   \qquad E_n^{(0)}=-{W_{nnnn}^2\over 16(n+1)^4}.$$
For some first $s$-states these equations give the following results
  $$ E_0^{(0)}=-0.097656\alpha^2,
    \qquad E_1^{(0)}=-0.0226\;\alpha^2, \qquad E_2^{(0)}=-0.0099\;\alpha^2,$$
which are different from the exact ones approximately by 10\%. 

Equation (11) is a special case of the following generalized equation
(see refs. [9,12])
  $$\Bigl(\hat A-E_n\hat B\Bigr)\vec{\Psi_n}=0. \eqno(17)$$
Therefore, we describe here a general iteration scheme for the last equation.
We shall find the eigenfunction in the form
  $$\vec{\Psi_n}=\vec{n}+\sum_{k\not=n}c_{kn}\vec{k}. \eqno(18)$$
The substitution of this expression into equation (17) gives the system of
equations ($s$ is a number of iterations)
  $$E^{(s)}_n=\Bigl(B_{nn}+\sum_{k\not=n}c^{(s)}_{kn}B_{nk}\Bigr)^{-1}
    \Bigl(A_{nn}+\sum_{k\not=n}c^{(s)}_{kn}A_{nk}\Bigr), \eqno(19)$$
  $$c^{(s)}_{in}=-\Bigl(A_{ii}-E^{(s-1)}_nB_{ii}\Bigr)^{-1}
    \sum_{k\not=n,i}\Bigl(A_{ik}-E^{(s-1)}_nB_{ik}\Bigr)c^{(s-1)}_{kn}, 
    \eqno(20)$$
where the eigenvalue $E_n$ and the coefficients of the eigenfunctions 
$c_{kn}$ are defined by the relations
  $$E_n=\lim_{s\to\infty} E_n^{(s)}, \qquad
    c_{kn}=\lim_{s\to\infty} c_{kn}^{(s)}. \eqno(21)$$
Equations (18)-(21) completely define quite a simple algorithm for solving
equation (17). Apparently, the non-diagonal matrix elements of the
Hamiltonian are ``included'' in each step of the subsequent iteration. It should be noted
that there exist some other iteration schemes, for example, an interesting
variational-iterative one, proposed by Burrows and Core (for more details, see \cite
{burrows}). Here we use the above scheme (18)-(21) because of its 
simplicity and transparency. Moreover, this and some other similar schemes proved to be
well convergent in many applications of the OM (see [8,9,12] and references therein).

The results of our calculation show good convergence of the OM in quite a
wide range of variation of the parameter $\omega$. For $s=18$ we obtain the
following value for the polaron ground-state energy
  $$E_0=-0.1085128052\;\alpha^2.$$
This result is in good agreement with the numerical solution,
obtained by Miyake \cite {miyake} and confirmed by Adamowski {\it et al} 
\cite {adam}. 
It should be noted that good numerical results have been obtained recently by a
variational method, based on the coherent state representations \cite{chen}. We would
like to point out that the OM enables one also to find with high accuracy
the energy and eigenfunction of the excited states, which usually are obtained
with much more difficulty using other methods \cite{gab}-\cite{bala}. We present here the 
value of the energy for $2s$- and $3s$-states (for $s=18$)
  $$E_1=-0.02053101\;\alpha^2, \qquad E_2=-0.0083506\;\alpha^2.$$
Our results are displayed in table 1 together with the known results for a 
polaron in the strong coupling limit.

\vskip 0.5cm
 {\bf Table 1.} Polaron energy (in $ \alpha^2 \hbar \omega_0$ 
units) for $s$-states in the strong coupling limit.\\

\begin{tabular}{llll}
\hline\hline
\quad\qquad Authors & $\qquad\quad 1s$ & $\qquad \  2s$ & $\qquad 3s$ \\
\hline
Pekar (by Miyake \cite {miyake})& $-0.108504$ &&\\
Miyake \cite {miyake}& $-0.108513$ &&\\
Adamowski {\it et al} \cite {adam}& $-0.1085128$ &&\\
Feranchuk {\it et al} \cite {2}& $-0.1078$ &&\\
Smondyrev \cite {smond1}& $-0.109206$ &&\\
Efimov {\it et al} \cite{efimov}& $-0.10843$ &&\\
Ganbold {\it et al} \cite{efimov2}& $ -0.107766$ &&\\
Chen {\it et al} \cite {chen} & $-0.10851$ &&\\
Hagen {\it et al} \cite {hagen}& & $-0.02048$ &$-0.00804$\\
Balabaev {\it et al} \cite {bala}& &$-0.0206$ &$-0.00832$\\
Our results  & $-0.1085128052$ & $-0.02053101$ & $-0.0083506$ \\
\hline \hline
\end{tabular}
\section{The operator method in solving the Hartree-Fock equation for helium}

Let us now consider the problem of helium. For the ground state, we have the
following Hartree-Fock equation (see, for example, ref. \cite {bethe})
  $$\left [-{1\over 2}\triangle_1-E_1-{Z\over r_1}+\int{\Psi^2(\rbf')\over
    |\rbf_1-\rbf'|}d\rbf'\right ]\Psi(r_1)=0 \eqno(22)$$
(for the second electron, the corresponding equation is written in analogy
with the first).

The energy of the atom is
  $$E=2E_1-\int\int{\Psi^2(\rbf_1)\Psi^2(\rbf_2)\over|\rbf_1-\rbf_2|}
    d\rbf_1d\rbf_2.$$
The substitution of the last expression into equation (22) gives
  $$\left [-{1\over 2}\triangle-{Z\over r}+\int{\Psi^2(\rbf')\over
    |\rbf-\rbf'|}d\rbf'-{1\over 2}
    \int\int{\Psi^2(\rbf_1)\Psi^2(\rbf_2)\over|\rbf_1-\rbf_2|}
    d\rbf_1d\rbf_2-{E\over 2}\right ]\Psi(r)=0. \eqno(23)$$
In $\xi$-space the equation (23) has the form
  $$\Biggl\{-{1\over 2}r\triangle -Z+{1\over\rsr}\int d^4\eta\;
    \Psi^2(\eta)[r'\theta(r-r')+r\theta(r'-r)]-$$
  $$-{r\over 2}\biggl({1\over \rsr^2}\int d^4\eta\int d^4\zeta\;
    \Psi^2(\eta)\Psi^2(\zeta)[r'\theta(r''-r')+r''\theta(r'-r'')]+E\biggr)
    \Biggr\}\Psi(\xi)=0. \eqno(24)$$
The last equation has the same form as equation (11), for which we have already 
calculated
the matrix elements. The key difference is that now we have a negative "coupling
constant" $\alpha$. In order to have the needed solutions, we
need only to change the formulae of the iteration scheme. Consequently,
for $s=18$ we have (in atomic units)
  $$E^{HF}=-2.861679995$$
which should be compared with the known result \cite{bethe}  $E^{HF}=-2.863$ 
and with a recent calculation \cite{sk} $E^{HF}=-2.861680$.
It follows that the OM gives the solutions with very high accuracy
independently of the repulsive or attractive nature of the interaction.

\section{Conclusions}
In this paper we have applied an algebraic approach, namely, the OM to solve the
non-linear equations, which appear
in the problem of polaron in the strong coupling limit and the helium atom in the
Hartree-Fock approximation. The resuls are as good as the best ones in the
literature. Furthermore, we would like to point out that the present approach should be
successfully applied to the other polaron problems, such as bound polarons and polarons in
external fields in 3D-space as well as in the other dimensions.

Finally, we note the following two circumstances:

(i) As shown in ref. \cite {kom84}, the equation, which describes the motion of
an individual particle in a system of large number of gravitational bosons in 
fact coincides with equation (11). Therefore, the solutions obtained here for
the polaron are simultaneously solutions for the above-mentioned problem.

(ii) We need not here only consider $s$-states. To this effect it is conducive
to use in equation (10) the following integral representation
  $${1\over |\rbf-\rbf'|}={1\over 2\pi}\int_0^{2\pi} d\chi 
    {1\over (\xi_s^*-\eta_s^*)(\xi_s-\eta_s)}.$$
Further calculation of matrix elements using this formula and the
algebra $so(4,2)$ of operators (3) does not pose
much difficulty and is affected in analogy with the calculations carried out
above.

\end{document}